\documentstyle[12pt,epsfig]{article}                                
\begin{document}

\titlepage                  

\begin{center}
\begin{Large}
\begin{bf}

Some studies  on : quantum equivalents of  non-commutative operators via commutating  eigenvalue relation: PT-symmetry 

\vspace{1.0cm}

\end{bf}
\end{Large}
\end{center}

Biswanath Rath

\vspace{0.1cm}

\begin{it}
 Department of Physics,
 Maharaja Sriram Chandra Bhanj Deo University,
 Takatpur, Baripada -757003, Odisha, India.
e.mail:biswanathrath10@gmail.com
\end{it}

\begin{large}

$\bf{Abstract:}$

Any  matrix "$B$" satisfying the non-commuting  relation $[A,B]\neq 0$ with "$A$", can be used via  $B^{-1}AB$ to reproduce eigenvalues of "$A$". 
This universality relation is also equally valid for any matrix in any branch of physical or social science and  also any  operator involving co-ordinate$(x)$ or
 momentum$(p)$. Pictorially this is represented  in the following fig. 
\end{large}
\begin{table}[htbp]
\begin{tabular}{ c  c } \\
 Matrix A  & Matrix $B^{-1}AB$ \\
\includegraphics[width=.5\textwidth]{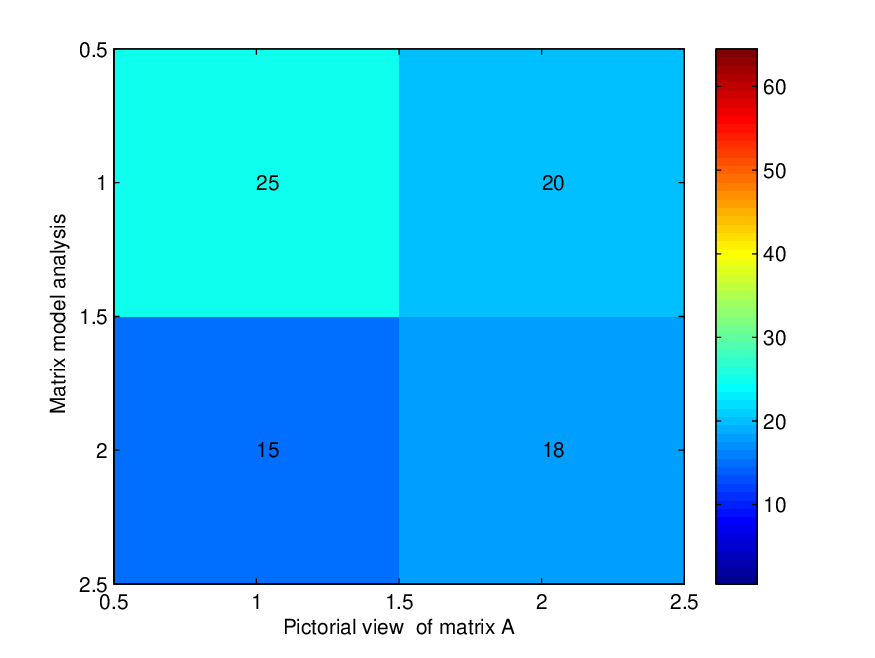} & 
\includegraphics[width=.5\textwidth]{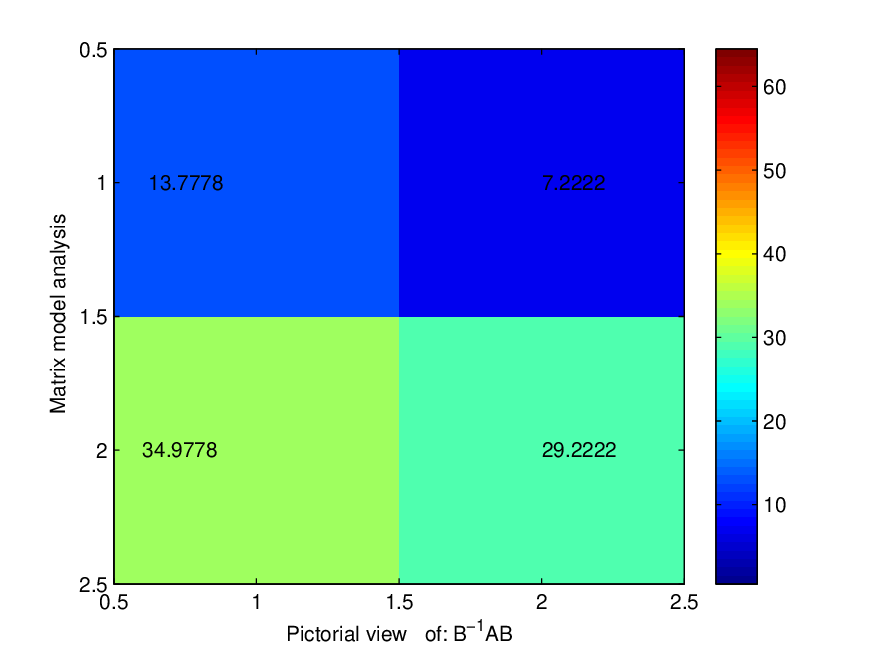} \\
    Eigenvalues of A &    Eigenvalues of $B^{-1}AB$ \\
\includegraphics[width=.5\textwidth]{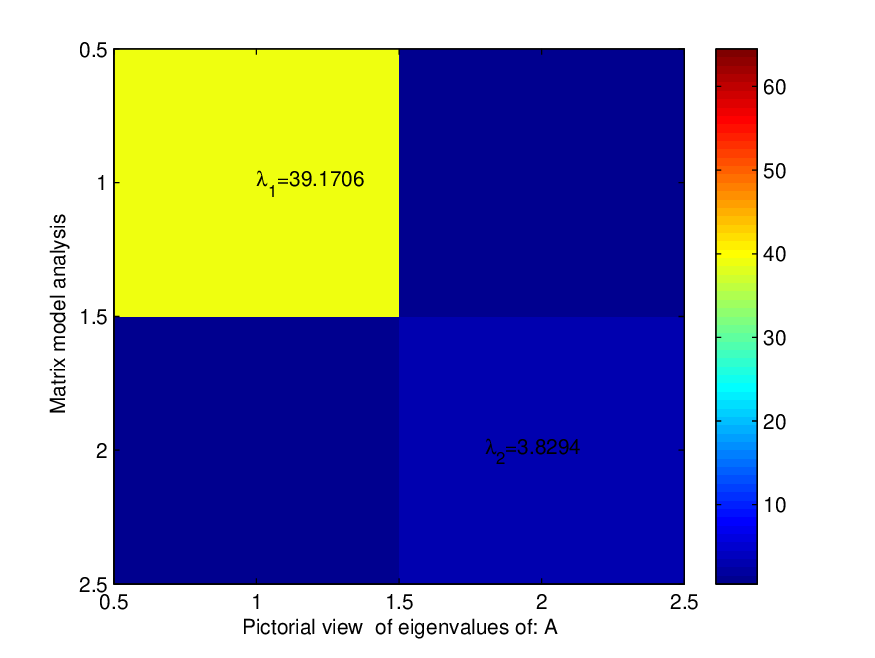} & 
\includegraphics[width=.5\textwidth]{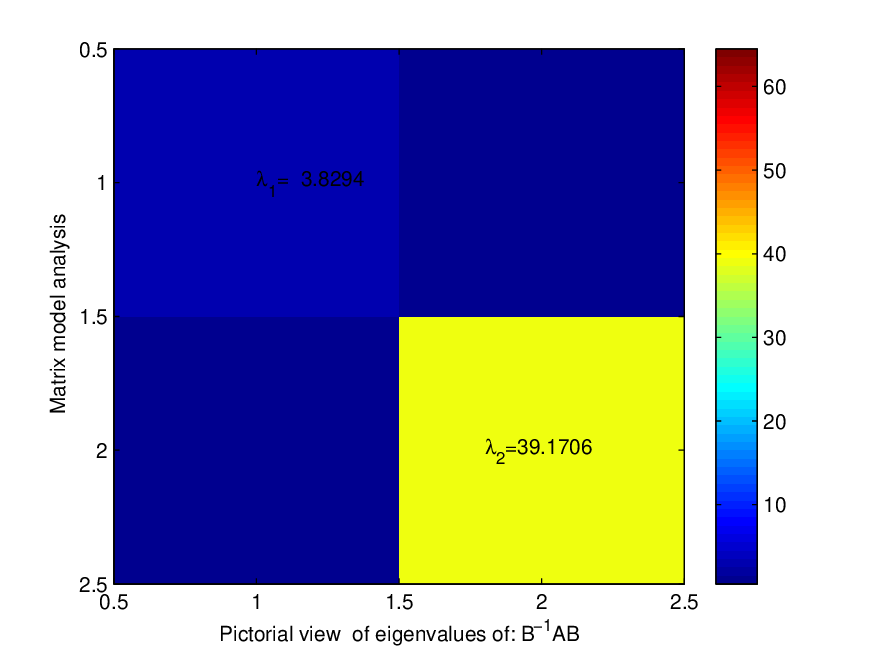} \\
\hspace{1.0in} FIG.1. Pictorial view  \\
\end{tabular}
\end{table}

Many interesting models including logarithmic potential have been considered.

\begin{bf}
PACS no-11.30Pb, 03.65.Ge
\end{bf}

\begin{bf}
\end{bf}

\pagebreak

\begin{bf}
I. Introduction
\end{bf}

It is commonly known as any two matrices having the same dimension and 
non-singular in nature are basically non-commutative in nature[1],

\begin{equation}
 AB \neq BA
\end{equation}
 Now we consider an explicit form of A or  B as 

\begin{bf}
Non-commutativity of matrices[1]
\end{bf}

 For example we consider two matrices as $A$ 

\begin{equation}
A =
\left[{\begin{array}{ c c }
 25 &  20 \\
15 &  18 \\
\end{array}}\right]
\end{equation}
having eigenvalues $\lambda_{1,2}=39.1706 ; 3.8294 $ and $B$ 

\begin{equation}
B =
\left[{\begin{array}{ c c }
8 &  10 \\
13 & 5 \\
\end{array}}\right]
\end{equation}
Here we do not pay importance to eigenvalues of $B$ ( whether reeal or complex)
.
However, $B$ must be non-singular in nature. Then it is easy to show that .

\begin{equation}
 AB \neq BA
\end{equation}

For a pictorial representation we consider 

\begin{equation}
\frac{AB}{10} =
\left[{\begin{array}{ c c }
46 &  35 \\
35.4 & 24 \\
\end{array}}\right]
\end{equation}

and

\begin{equation}
\frac{BA}{10} =
\left[{\begin{array}{ c c }
35 &  34 \\
40 &  35 \\
\end{array}}\right]
\end{equation}

Pictorially this is represented as in Fig-1.

\begin{table}[htbp]
\begin{tabular}{ c  c } \\
\includegraphics[width=.5\textwidth]{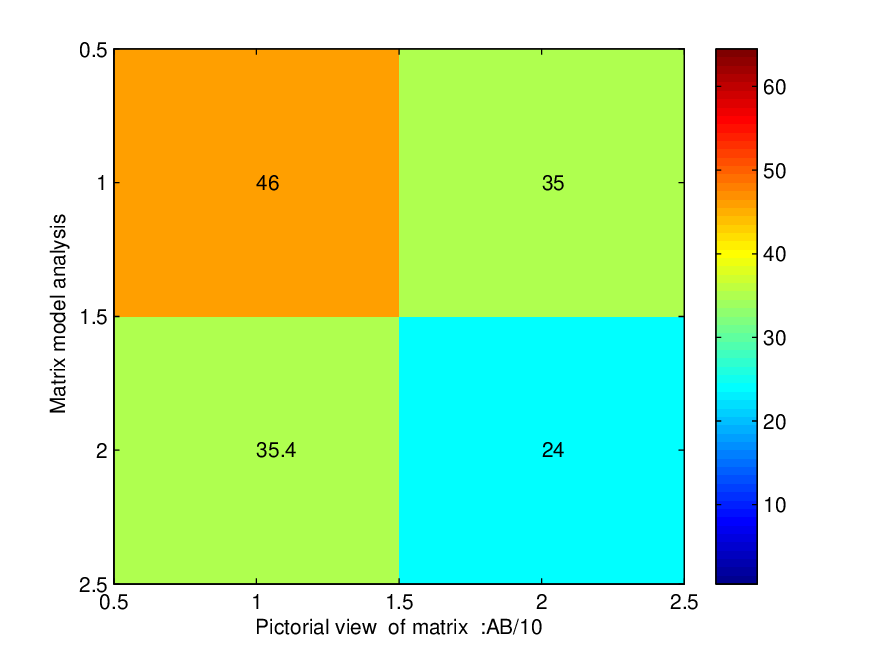} & 
\includegraphics[width=.5\textwidth]{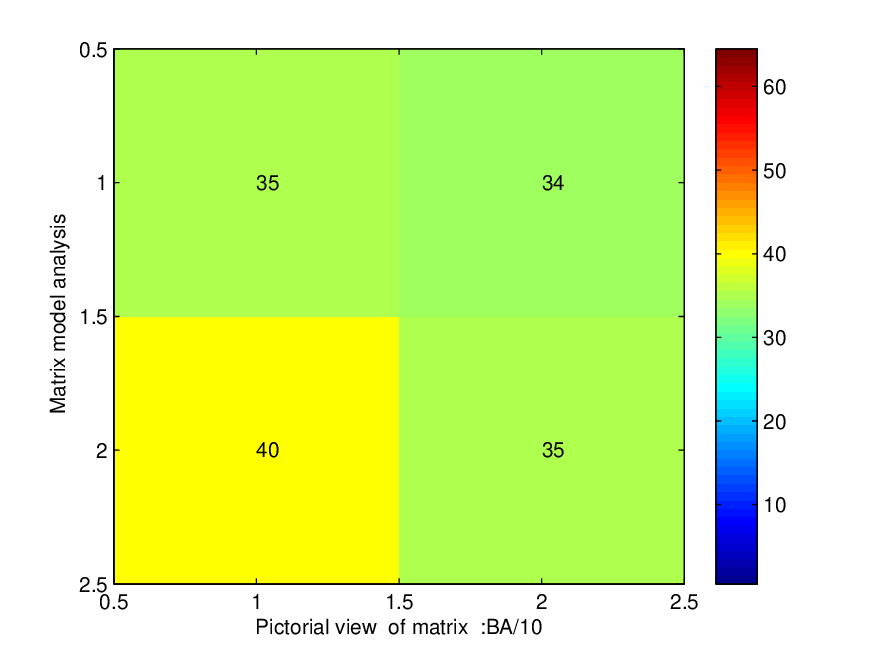} \\
\hspace{1.0in} FIG.1.Non-commutativity nature  \\
\end{tabular}
\end{table}

As stated above, we have the freedom to change $B$ as $B_{k}$, where $k=1,2,3,4,.........$ . Below we consider a few different cases as 

\begin{bf}
Hermiticity   
\end{bf}

Here we consider $B_{1}$ as a Hermitian operator as 

\begin{equation}
B_{1} =
\left[{\begin{array}{ c c }
8  &  10+3i  \\
10-3i & 8     \\
\end{array}}\right]
\end{equation}
 The corresponding, transformed matrix is  

\begin{equation}
B_{1}^{-1} A B_{1} =
\left[{\begin{array}{ c c }
-0.8444 + 18.6667 i & -10.5556 + 16.2667 i  \\
31.5556 -30.4000 i & 43.8444 - 18.6667 i     \\
\end{array}}\right]
\end{equation}

whose eigenvalues are the same as $A$.

\begin{bf}
PT-symmetry[2] 
\end{bf}

Here the model matrix satisfies the condition $[B_{2},PT]=0$. 
\begin{equation}
B_{2} =
\left[{\begin{array}{ c c }
8 \pm 2i &  10 \\
10 & 8 \mp 2i  \\
\end{array}}\right]
\end{equation}

.Here $P$ stands for the parity operator having the nature :$P x P^{-1}=-x$; $P p P^{-1}=-p$. Similarly T stands for the time reversal operator having the behaviour: $T x T^{-1}=x $; $T p T^{-1}= -p$ and $T i = -i$[2].. The model operators, which have been studied for long time by different authors[2-5]
\begin{equation}
B_{2}^{-1} A B_{2} =
\left[{\begin{array}{ c c }
-9.3750 + 21.8750 i & -8.1250 + 24.3750 i  \\
51.8750 -10.6250 i &  52.3750 - 21.8750 i     \\
\end{array}}\right]
\end{equation}

Here also , the eigenvalues are the same as A

\begin{bf}
Arbitrary non-singular matrix 
\end{bf}

Here the the matrix considered is very arbitrary in nature.For example 
\begin{equation}
B_{3} =
\left[{\begin{array}{ c c }
8 \pm 2i &  10 +i \\
10 & 8   \\
\end{array}}\right]
\end{equation}

\begin{equation}
B_{3}^{-1}AB_{3} =
\left[{\begin{array}{ c c }
-9.3750 + 21.8750 i  & - 8.1250 + 24.3750 i  \\
51.8750 - & 8     \\
\end{array}}\right]
\end{equation}

In all the cases, we find that eigenvalues remain the same as A.

\begin{bf}
2. Universality of eigenvalue invariance (via commutative relation)
\end{bf}

Let us confine to eigenvalues of $A=H$, using the relation 

\begin{equation}
B_{k}= h_{k} H h_{k}^{-1}
\end{equation}

Here the relation to be read as eigenvalues of LHS = eigenvalues of RHS. Mathematically for eigenvalue determination, it is basically a commutative relation (even though,non-commutative in general).

Below we confine our analysis on complex PT-symmetry operators in matrix as well as operator forms as given below..

\begin{bf}
3. Spectral invariance in PT-symmetry operator 
\end{bf}

 Let us confine our attention to $A=H^{PT}=H$ as a PT-symmetry operator in matrix model.

 Here, we consider the  Hamiltonian(H) as  (2x2) matrix  model as

\begin{bf}
First model
\end{bf}

\begin{equation}
H =
\left[{\begin{array}{ c c }
 1+\sqrt{3} i & 4 \\
4  &  1-\sqrt{3} i \\
\end{array}}\right]
\end{equation}
The above model has energy levels $\lambda_{1}=4.6056 $ and $\lambda_{2}=-2.6056 $.
Let us choose  different forms of  $B_{k}$ as 

\begin{equation}
h_{1} =
\left[{\begin{array}{ c c }
1  & 3 + 4i \\
3-4i  & 1 \\
\end{array}}\right]
\end{equation}

 The corresponding $B_{1}$ becomes
\begin{equation}
B_{1} =
\left[{\begin{array}{ c c }
1- 3.2097 i & -1.9107 + 4.4330 i \\
-1.9107 - 4.4330 i&  1 + 3.2097 i  \\
\end{array}}\right]
\end{equation}

\begin{bf}
Second model
\end{bf}

 Here we consider  $h_{2}$ as 

\begin{equation}
h_{2} =
\left[{\begin{array}{ c c }
2  & 3 + 4i \\
3-4i  & 2 \\
\end{array}}\right]
\end{equation}

The corresponding $B_{2}$ becomes

\begin{equation}
B_{2} =
\left[{\begin{array}{ c c }
1 - 5.4395 i & -3.4149 + 5.5612 i \\
-3.4149 - 5.5612 i & 1=5.4395 i \\
\end{array}}\right]
\end{equation}

\begin{bf}
Third  model
\end{bf}

 Here we consider  $h_{3}$ as 

\begin{equation}
h_{3} =
\left[{\begin{array}{ c c }
3  & 3 + 4i \\
3-4i  & 3 \\
\end{array}}\right]
\end{equation}

The corresponding $B_{3}$ becomes 

\begin{equation}
B_{3} =
\left[{\begin{array}{ c c }
1-9.6806 i  & -6.5981 + 7.9486 i \\
-6.5981 - 7.9486 i  & 1+ 9.6806  i \\
\end{array}}\right]
\end{equation}

\begin{bf}
Fourth   model
\end{bf}

 Here we consider $h_{4}$ as 

\begin{equation}
h_{4} =
\left[{\begin{array}{ c c }
 4  &  1 + 3i \\
1-3i  & 4  \\
\end{array}}\right]
\end{equation}
The corresponding $h_{4}$ becomes

\begin{equation}
B_{4} =
\left[{\begin{array}{ c c }
1-22.1127 i & -16.3806 + 15.2855 i \\
-16.3806 -15.2855 i  &  1- 22.1127 i \\
\end{array}}\right]
\end{equation}

In all the cases, eigenvalues of $B_{k}$ remain the same as $H$. Pictorially, we have 

\begin{table}[htbp]
\begin{tabular}{  c } \\
\includegraphics[width=.5\textwidth]{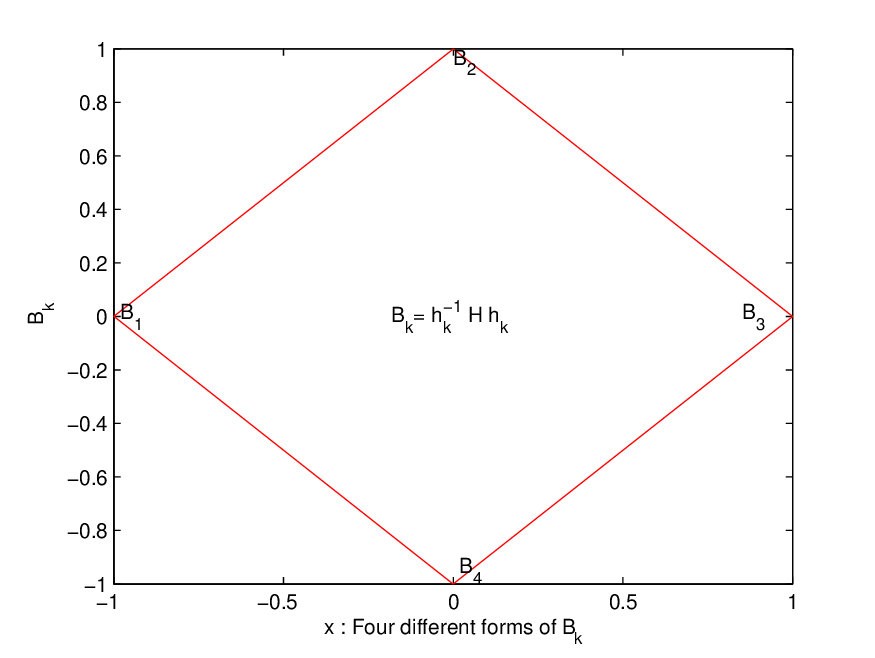} \\
\hspace{1.0in} FIG.2.Equi-eigenvalue model operators  \\
\end{tabular}
\end{table}

\pagebreak

\begin{bf}
4a. Operator model involving $x$ and $p$.
\end{bf}

Here we consider diffent operators as follows.

\begin{bf}
 Hermitian : $ h_{1} + p^{2} + x^{2} $
\end{bf}

Let us consider a simple harmonic oscillator as $h_{1}$ i.e

\begin{equation}
h_{1}=p^{2}+ x^{2}
\end{equation}

whose energy levels are real and equispaced.

\begin{equation}
E^{1}_{n} = (2n+1)
\end{equation}

Now consider another hermitian operator as 

\begin{equation}
h_{2}=p^{2}+ x^{4}
\end{equation}

having real spectra [3]. As stated above $h_{k}$ can also ave broken spectra.
Accordingly, we consider a broken spectra  $h_{4}$ as

\begin{equation}
h_{3}=p^{2}+ ix 
\end{equation}

whose spectra is complex see fig-3.

\begin{bf}
\begin{table}[htbp]
\begin{tabular}{  c } \\
\includegraphics[width=.5\textwidth]{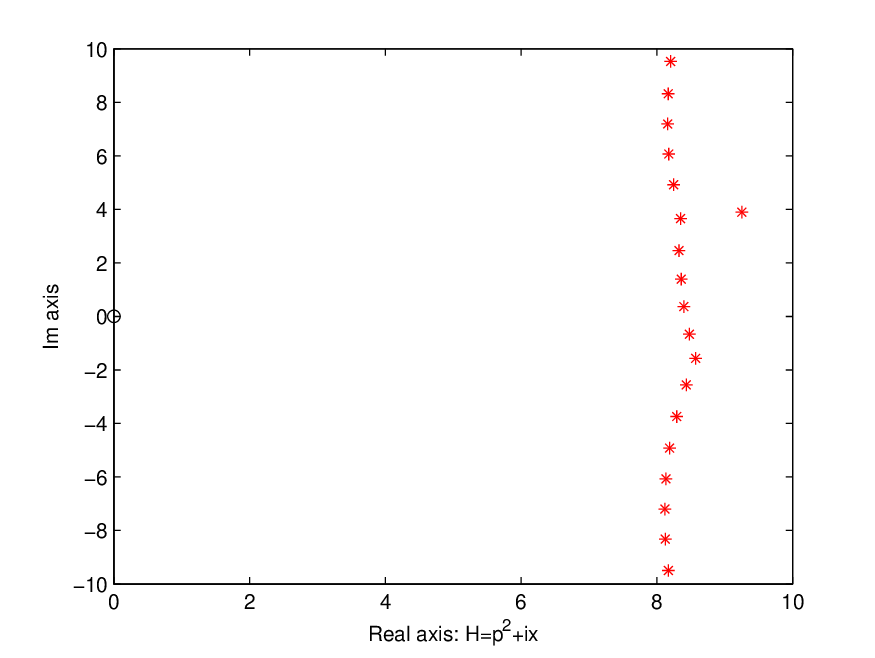} \\
\hspace{1.0in} FIG3: Spectra of $H=p^{2} +ix$   \\
\end{tabular}
\end{table}
\end{bf}

Similarly , we also consider another broken PT-symmetry operator as[5] 

\begin{equation}
h_{4}=p^{2}+ \frac{i}{x}
\end{equation}

Now use the flowing  PT symmetry operator as

\vspace{0.1cm}

\vspace{0.1cm}
\begin{table}[htbp]
\begin{tabular}{  c } \\
\includegraphics[width=.5\textwidth]{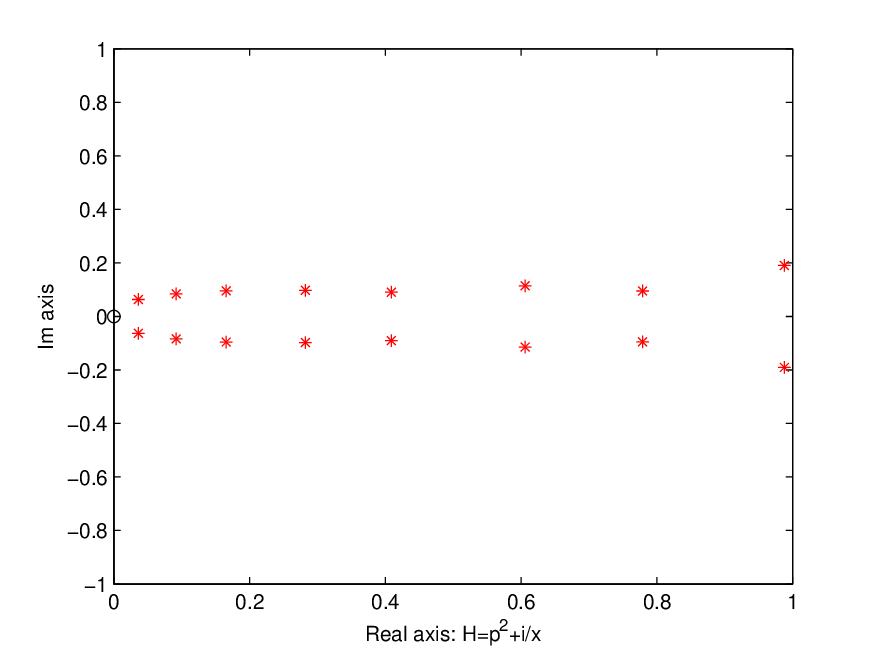} \\
\hspace{1.0in} FIG.2.Spectra of $H=p^{2}+\frac{i}{x}$    \\
\end{tabular}
\end{table}

\pagebreak

The corresponding $B$ operator  is 
\begin{equation}
B_{k}= h_{k} H_{2} h_{k}^{-1}
\end{equation}
Let us consider $H_{2}$ as[4,5]
\begin{equation}
H_{2}= p^{2} + ix^{3}
\end{equation}

The energy levels are tabulated in table-1.

\begin{center}

\begin{table}[htbp]

 Table-1: First five even  energy levels of $B_{k}$  \\

 \begin{tabular}{ c c  c   } \\ 
        $B_{1} $ & $B_{2}$ & $B_{3}$ \\ \hline \hline
        1.156 267  &  1.156 267   & 1.156 267  \\
        7.562 273  &  7.562 273  &  7.562 267  \\
        15.291 553 & 15.191 553 & 15.291 553  \\
        23.766 740 & 23.766 740 & 23.766 740 \\ 
        32.789 082 & 32.766 082 & 32.789 082 \\ \hline
       $B_{4} $ & Previous[7] & Previous[4] \\ \hline 
       1.156 267  &  1.156 2   & 1.156 267  \\
        7.562 273  &  7.562 1  &  7.562 267  \\
        15.291 553 & 15.291 6  & 15.291 553  \\
        23.766 740 &        & 23.766 740 \\  \hline \hline
\end{tabular}
\end{table}
\end{center}

\begin{bf}
4b. Inverted quartic operator model[5-7] involving $x$ and $p$.
\end{bf}

\vspace{0.1cm}

$ H_{3} = p^{2} - x^{4} $

\vspace{0.1cm}

The corresponding $B$ operator  is 

\begin{equation}
B_{k=1,2,3,4}= h_{1,2,3,4} H_{3} h_{1,2,3,4}^{-1}
\end{equation}

\begin{center}

\begin{table}[htbp]

\vspace{1.0cm}

 Table-2: First five even  energy levels of $B_{k}$  \\

      \begin{tabular}{ c c  c   } \\ 

        $B_{1} $ & $B_{2}$ & $B_{3}$ \\ \hline \hline
        1.477 149 7   & 1.477 149 7  & 1.477 149 7  \\
        11.802 433 5   &11.802 433 5 & 11.802 433 5 \\
        25.791 792 3  & 25.791 792 3 & 25.791 792 3  \\
        42.093 807 7 &  42.093 807 7 & 42.093 807 7 \\ 
       60.184 331 2 &  60.184 331 2 & 60.184 331 2 \\ \hline
  $B_{4} $ & exact[5,6] & Previous[4] \\ \hline 
        1.477 149 7   & 1.477 149 7  & 1.477 149 7  \\
        11.802 433 5   &11.802 433 5 & 11.802 433 5 \\
        25.791 792 3  & 25.791 792 3 & 25.791 792 4  \\
       42.093 807 7 &  42.093 807 7 & 42.093 814 5 \\ 
        60.184 331 2 &  60.184 331 2 & 60.185 767 6 \\ \hline \hline

\end{tabular}
\end{table}
\end{center}

\pagebreak

\begin{bf}
4c.Logarithmic model  quartic PT-symmetry[8]  involving $x$ and $p$.
\end{bf}

\vspace{0.1cm}

$ H_{4} = p^{2} + x^{4}\log(ix) $

\vspace{0.1cm}

The corresponding $B$ operator  is 
\begin{equation}
B_{k=1,2,3,4}= h_{1,2,3,4} H_{4} h_{1,2,3,4}^{-1}
\end{equation}

Here, we present results in table3.

\begin{center}

\begin{table}[htbp]

\vspace{1.0cm}

 Table-3:  Energy levels of $B_{k}$  \\

      \begin{tabular}{ c c  c   } \\ 

        Quantum no & $B_{1}$ & $B_{2}$ \\ \hline \hline
         0 & 1.249 08  &  1.249 08    \\
         3 & 13.738 27 & 13.738 27  \\
         6 & 31.665 82 & 31.665 82 \\
         9 & 52.993 79 & 52.993 79 \\
         12& 76.976 08 & 76.976 08   \\ \hline
  $B_{3} $ & $B_{4}$ & Previous[6] \\ \hline 
         0 & 1.249 08  &  1.249 08    \\
         3 & 13.738 27 & 13.738 27  \\
         6 & 31.665 82 & 31.665 82 \\
         9 & 52.993 79 & 52.993 79 \\
         12& 76.976 08 & 76.976 08   \\ \hline \hline

\end{tabular}
\end{table}
\end{center}

\pagebreak

\begin{bf}
5.Method of calculation
\end{bf}

Here we solve the eigenvalue relation[9-11] 

\begin{equation}
H|\Psi>=E|\Psi>
\end{equation}
where
\begin{equation}
|\Psi>=\sum A_{m} |m>
\end{equation}
where $|m>$ satisfies the eigenvalue relation[8]
\begin{equation}
[p^{2}+x^{2}]|m>=(2m+1)|m>
\end{equation}.

\begin{bf}
6.Conclusion
\end{bf}

Here, we have focussed attention on PT-symmetry model both matrix and operator 
involving$ (x,p)$ and presented a unique spectral of Hamiltonian under the 
transformation. In fact, " intertwining" operators $B_{k}; h_{k};H$. It should be borne in mind that the operators$h_{k}$ be non-singular. The eigenvalues of $h_{k}$ are not an important criteria on selection of $B_{k}$. It can have  broken spectra also. To justify this, we have considered two potentials $V_{1}=ix$ 
and $V_{2}=\frac{i}{x}$.
Lastly one can have many more operators like this, in which no specific conditions to be imposed on $h_{k}$.

\begin{bf}
Conflict of interest:
\end{bf}

Author declares there are no conflicts of interest.

\begin{bf}
DATA Requirement : No additional data of any kind is required.
\end{bf}

\end{document}